\documentclass[12pt,a4paper]{article}
\usepackage{graphicx}
\usepackage{dcolumn}
\usepackage{bm}
\usepackage{epstopdf}
\usepackage{amsmath}
\usepackage{tabularx}
\usepackage{float}
\usepackage{caption}
\usepackage{array}
\usepackage{amstext}
\usepackage{footnote}
\usepackage{footmisc}
\usepackage{setspace}
\usepackage{authblk}
\usepackage{pdfpages}

\usepackage[left=1.0in, right=1.in, top=0.95 in, bottom=0.95in]{geometry}

\begin{document}
\begin{center}
\doublespacing{\large{\textbf{Layered transition metal dichalcogenides: promising near-lattice-matched substrates for GaN growth}}}\\[0.5cm]

\onehalfspacing{\normalsize{
Priti Gupta,$^1$\footnote{Presently at Department of Materials Science and Metallurgy, University of Cambridge, Cambridge, UK}
A. A. Rahman,$^1$
Shruti Subramanian,$^1$\footnote{{Presently at Department of Materials Science and Engineering, Pennsylvania State University, Pennsylvania, USA}}
Shalini Gupta,$^{1,2}$
Arumugam Thamizhavel,$^1$
Tatyana Orlova,$^3$
Sergei Rouvimov,$^3$
Suresh Vishwanath,$^3$\footnote{\label{cornell}{Presently at Department of  Electrical and Computer Engineering, Department of Materials Science and  Engineering, Cornell University, Ithaca, NY, USA}}
Vladimir Protasenko,$^3$
Masihhur R. Laskar,$^4$
Huili Grace Xing,$^3$\footnotemark[\value{footnote}]
Debdeep Jena,$^3$\footnotemark[\value{footnote}] \\
Arnab Bhattacharya$^{1}$}}\footnote{Email: arnab@tifr.res.in}\\[0.3cm]

\textit{$^1$Department of Condensed Matter Physics and Materials Science, Tata Institute of Fundamental Research, Mumbai, India}\\
\textit{$^2$UM-DAE Center for Excellence in Basic Sciences, Mumbai, India}\\
\textit{$^3$Department of Electrical Engineering, University of Notre Dame, Notre Dame, USA}\\
\textit{$^4$Department of Chemical and Biological Engineering, University of \\Wisconsin-Madison, Madison, USA}

\end{center}

\vspace{0.5cm}

\begin{abstract}
\onehalfspacing{\small{Most III-nitride semiconductors are grown on non-lattice-matched substrates like sapphire or silicon due to the extreme difficulty of obtaining a native GaN substrate. We show that several layered transition-metal dichalcogenides are closely lattice matched to GaN and report the growth of GaN on a range of such layered materials. We report detailed studies of the growth of GaN on mechanically-exfoliated flakes WS$_2$ and MoS$_2$ by metalorganic vapour phase epitaxy. Structural and optical characterization show that strain-free, single-crystal islands of GaN are obtained on the underlying chalcogenide flakes. We obtain strong near-band-edge emission from these layers, and analyse their temperature-dependent photoluminescence properties. We also report a proof-of-concept demonstration of large-area epitaxial growth of GaN on CVD MoS$_2$. Our results show that the transition-metal dichalcogenides can serve as novel near-lattice-matched substrates for nitride growth. }}
\end{abstract}

\pagebreak

\doublespacing
\setlength{\parindent}{0pt}

\section{\large{Introduction}}
The group-III nitride semiconductors\cite{ambacher1998growth} --- GaN, AlN, InN and their alloys --- are important materials for compact, energy-efficient, solid-state lighting sources as well as ideal candidates for high-power\cite{baliga2013gallium} / high-temperature\cite{neudeck2002high} electronic devices. There have been phenomenal improvements in the performance of GaN-based LEDs,\cite{wierer2009iii,ponce1997nitride} laser diodes,\cite{arafin2013review} and transistors;\cite{chowdhury2013current} however, a fundamental challenge in these materials has been the extreme difficulty of obtaining a native GaN substrate compelling the use of non-lattice-matched substrates like sapphire or silicon.\cite{neumayer1996growth,gibart2004metal,kukushkin2008substrates} While breakthroughs\cite{nakamura1991gan,amano1986metalorganic} in epitaxial growth techniques have dramatically improved material quality on sapphire, there has always been a quest to seek lattice-matched substrate materials for GaN epitaxy. In this letter, we show that several layered transition metal dichalcogenides\cite{wang2012electronics} are closely lattice matched to GaN and report the metalorganic vapour phase epitaxial (MOVPE) growth of GaN on a range of such layered materials. In particular, we show WS$_2$ and MoS$_2$ to be promising substrates for III-nitride growth.

\begin{figure}[ht]
\centerline{\includegraphics[width=88mm]{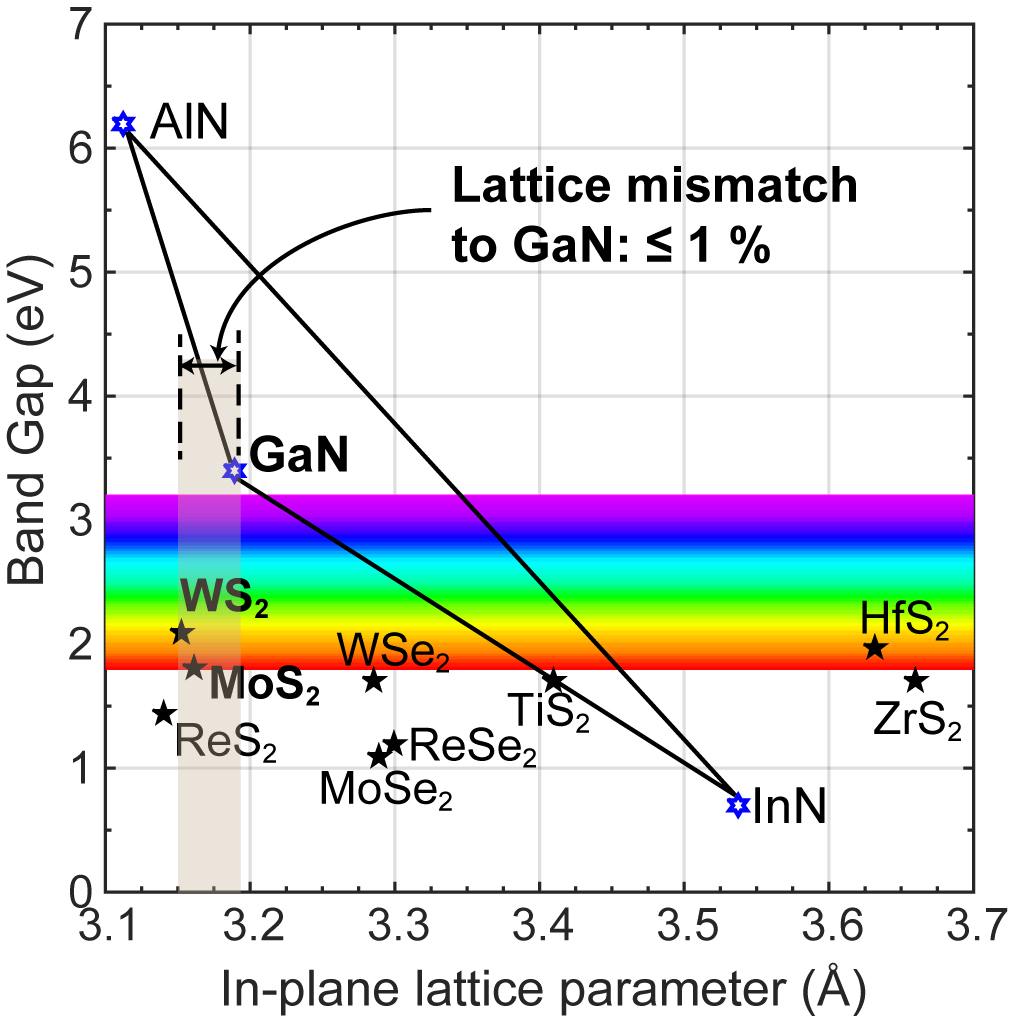}}
\caption{\onehalfspacing{\textbf{Band gap versus in-plane lattice parameter for different III-nitrides and TMDCs.} The lattice mismatch of WS$_2$ and MoS$_2$ with respect to GaN are 1.0~$\%$ and 0.8~$\%$, respectively. Supplementary information section I lists the sources from where the lattice parameters and bandgaps of different III-nitrides and TMDCs materials were obtained. }}
\label{fig1}
\end{figure}

The past few years have seen a frenzy of activity in research in 2-dimensional layered materials, motivated by the exceptional properties of graphene\cite{geim2007rise} and the demonstration of a range of novel devices. The transition-metal dichalcogenides (TMDCs)\cite{wang2012electronics} --- layered materials analogous to graphene but which also offer a bandgap\cite{mak2010atomically} ---  have been a subject of great recent interest due to their own unique electrical and optical properties. Most of the work in semiconducting layered materials has focused on their potential for opto-electronic devices\cite{roy2013graphene,lopez2013ultrasensitive,yin2011single,perea2013photosensor,radisavljevic2011single}. What has perhaps not been noticed is their relatively close in-plane lattice match with materials of the III-nitride family. Figure~\ref{fig1} shows the bandgap versus in-plane lattice parameter for different III-nitrides and TMDCs. WS$_2$ and MoS$_2$ have a lattice mismatch of only 1.0\% and 0.8\%, respectively  to the \textit{`a'} lattice parameter of GaN. Further, the weak out-of-plane (van der Waal's) interactions and absence of dangling bonds on the surface of the layered chalcogenides may help in controlling stress between the nitride epilayer and substrate. The TMDCs can thus not only provide novel near-lattice-matched substrates for nitride growth, but are also likely to result in new materials combinations and heterostructures  enabling flexibility in the design of optoelectronic devices.

\section{\large{Results and discussion}}

To test their viability as substrates for nitride growth, mechanically-exfoliated flakes of the layered TMDCs were transferred to SiO$_2$ (300~nm thick) coated silicon wafers (details in supplementary information section II).  We chose SiO$_2$-coated Si rather than sapphire, Si or SiC to minimize parasitic growth outside the flakes which may hinder the characterization of the GaN layer. In addition, for a proof-of-concept demonstration of large-area epitaxial growth, CVD MoS$_2$ films synthesized by the sulphurization of thin Mo layers were used (details in supplementary information section II).

The heat-up procedure, carrier gas, nucleation conditions, growth temperature and V/III ratio were varied to determine the optimal conditions for growth of GaN on WS$_2$ and MoS$_2$. The MOVPE growth of GaN typically occurs at a high temperature, $>$~1000~$^o$C, in an ambient of ammonia and hydrogen, rather aggressive conditions, which are in some cases close to the thermal decomposition temperatures of the chalcogenide materials. This makes the procedure for initiation of growth on the TMDC material very critical, especially the choice of carrier gas and temperature ramp during the heat-up procedure, and the conditions for the growth of the nucleation layer. Our observations suggest that ramping up to the growth temperature is best done using nitrogen (N$_2$) as a carrier gas to preserve the integrity of the TMDC layer.  We switch the carrier gas to hydrogen just before the growth of the GaN layer, and again cool down to room temperature under N$_2$. Based on optimization experiments, we have used a short initial GaN nucleation layer grown at 900~$^o$C for 40~s at a V/III ratio of $\sim$~1400, following which a second layer is grown at the more usual growth temperature of 1040~$^o$C at a V/III ratio of $\sim$~2100 (we will refer this second layer growth time as t$_{GaN}$). The low temperature GaN layer also ensures the conformal coverage of GaN on the TMDC substrate.

\subsection{\normalsize{Growth of GaN on WS$_2$}}

\begin{figure*}[p]
\centerline{\includegraphics[width=1.0\textwidth]{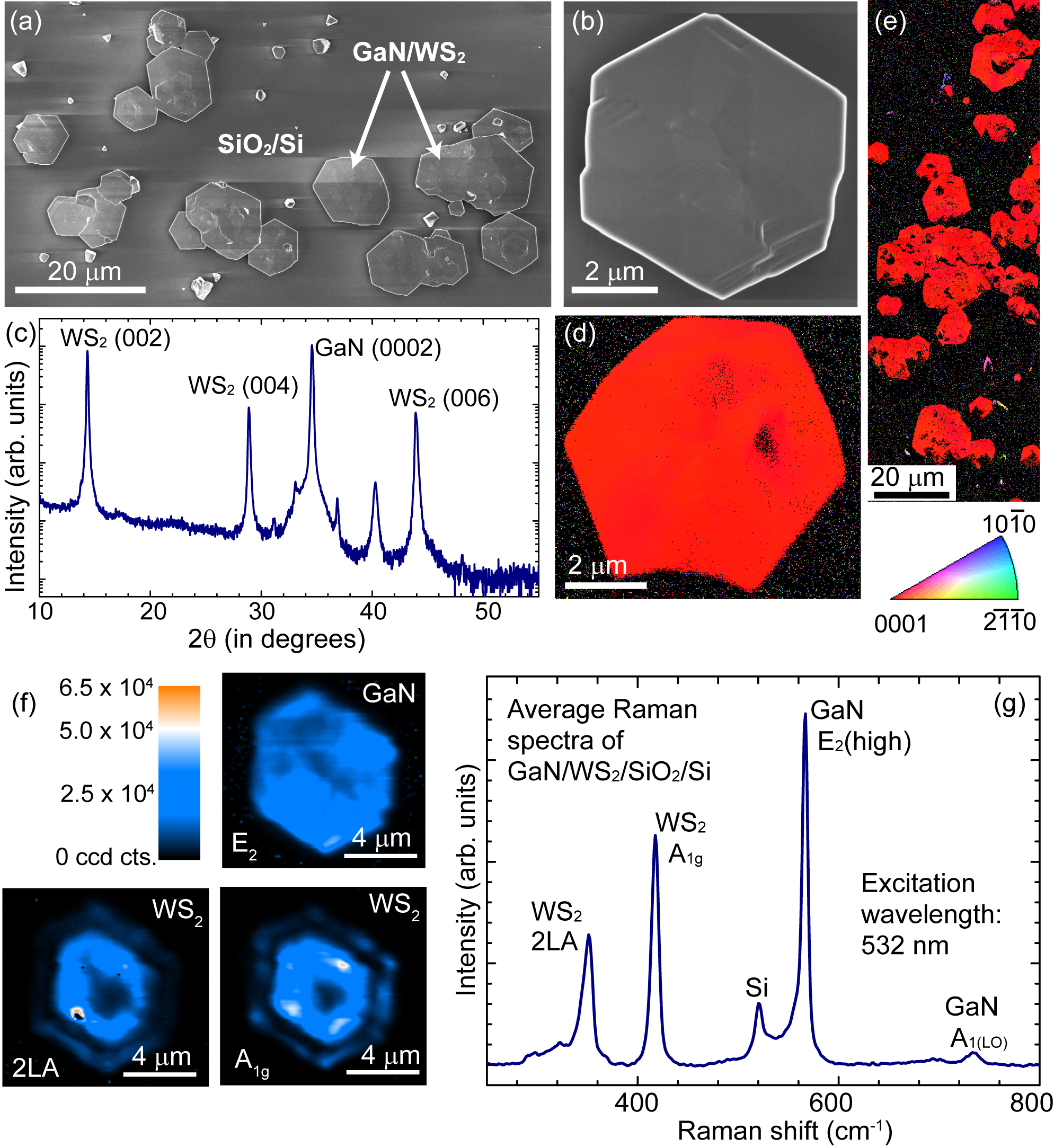}}
\caption{\onehalfspacing{\textbf{Growth of GaN on exfoliated WS$_2$ flakes.} (a) Scanning electron microscopy confirms near-hexagonal crystals of GaN growing only in the region covered by the WS$_2$ flakes. (b) Micrograph showing single hexagonal crystal of GaN grown on WS$_2$ (c) X-ray diffraction profile of GaN on WS$_2$ shows preferential (0002) orientation. (d) and (e) EBSD maps of GaN grown on exfoliated WS$_2$ clearly show that the epitaxial GaN layer is single crystal and (0002) oriented. (f) Integrated Raman mapping over an area of 10 $\mu$m $\times$ 11 $\mu$m for the intensity of following Raman modes E$_2$(high) of GaN, 2LA and A$_{1g}$ mode of WS$_2$. (g) Spatially averaged Raman scattering spectrum of GaN/WS$_2$ over the flake shown in (f) shows the survival of WS$_2$ after growth and the peak position of E$_2$(high) indicates that GaN epilayer on WS$_2$ is strain-free.}}
\label{fig2}
\end{figure*}

 We first discuss the growth of GaN on WS$_2$. For t$_{GaN}$= 300 s, nearly hexagonal crystals of GaN are obtained only on WS$_2$ flakes as shown in Figure~\ref{fig2}(a) and (b). From the XRD profile of GaN on WS$_2$ (Figure~\ref{fig2}(c)), it is clearly seen that GaN is growing in the (0002) direction as expected. The few stray peaks are possibly due to either random nucleation on SiO$_2$/Si or due to growth on mis-oriented WS$_2$ flakes stamped on the substrate. All the GaN/WS$_2$ flakes are oriented in (0002) direction as confirmed from the large area electron back scattering diffraction (EBSD) map (Figure~\ref{fig2}(e)). A detailed EBSD map of one flake (Figure~\ref{fig2}(d)), shows that the GaN is single crystal with no grain boundary. Similar EBSD maps on several other flakes confirm the single crystal growth of GaN on WS$_2$. The observation of a (0002) oriented GaN layer is not surprising. Even if there are no bonds across the interface, the atomic arrangement of the underlying layer would still influence the energy landscape on the growth surface and hence have an effect on the nucleation and growth of the epilayer. Thus, the lattice structure of the substrate will influence the epitaxial relation. For closely-lattice-matched systems, as in the case of the TMDCs, the substrate\textsc{\char13}s lattice points will be the energetically favoured regions and even if there is little strain arising from some mismatch, it is easily relaxed because of the absence of strong interlayer bonding. The lattice parameters of GaN were calculated from the powder XRD peaks; the obtained value of the in-plane lattice parameter was 3.188 $\AA$ which is almost equal to in-plane lattice parameter reported for unstrained GaN i.e 3.189 $\AA$. Thus, as expected, the GaN layer being grown on a closely-lattice matched substrate is almost strain-free.

Raman spectroscopy was used to examine the residual stress and to monitor the quality of the GaN epilayer. The frequency shift of the strongest phonon mode E$_2$(high) is very sensitive to the residual stress\cite{kisielowski1996strain,kitamura2008raman} whereas the width depends on the quality of the layer.\cite{kitamura2008raman} The lattice- and thermal-mismatch between the layer and the substrate result in a residual strain which is usually compressive for GaN grown on sapphire, and tensile for GaN on Si.\cite{zhao2003stress} The E$_2$(high) mode of strain-free GaN is known to be at 566.2 cm$^{-1}$ at room temperature,\cite{kisielowski1996strain} and shifts to higher frequency with compressive strain whereas a downshift indicates tensile strain. Figure~\ref{fig2}(f) shows the Raman integrated maps of the E$_2$(high) mode of GaN, and 2LA and A$_{1g}$ modes of WS$_2$. The homogeneous intensity distribution of the E$_2$(high) peak also points to the uniform GaN growth over the WS$_2$ flake. The spatially-averaged Raman spectrum of GaN/WS$_2$ over a 10 $\rm\mu m \times$ 11 $\rm\mu m$ area is shown in Figure~\ref{fig2}(g).  The spatial extent of the GaN may be larger than the underlying WS$_2$ flake due to lateral overgrowth at the edges. The E$_2$ phonon peak for GaN/WS$_2$ is observed at 566.4 cm$^{-1}$ which is almost near the value reported for the unstrained GaN. This also confirms that the GaN epilayer grown on WS$_2$ is nearly strain-free. The linewidth of the E$_2$(high) peak is 6~cm$^{-1}$, comparable with the linewidth of the E$_2$(high) Raman peak of (0002)-oriented GaN of similar thickness grown on sapphire with the standard two-temperature recipe.

\subsection{\normalsize{Growth of GaN on MoS$_2$}}

\begin{figure*}[h!]
\centerline{\includegraphics[width=1.0\textwidth]{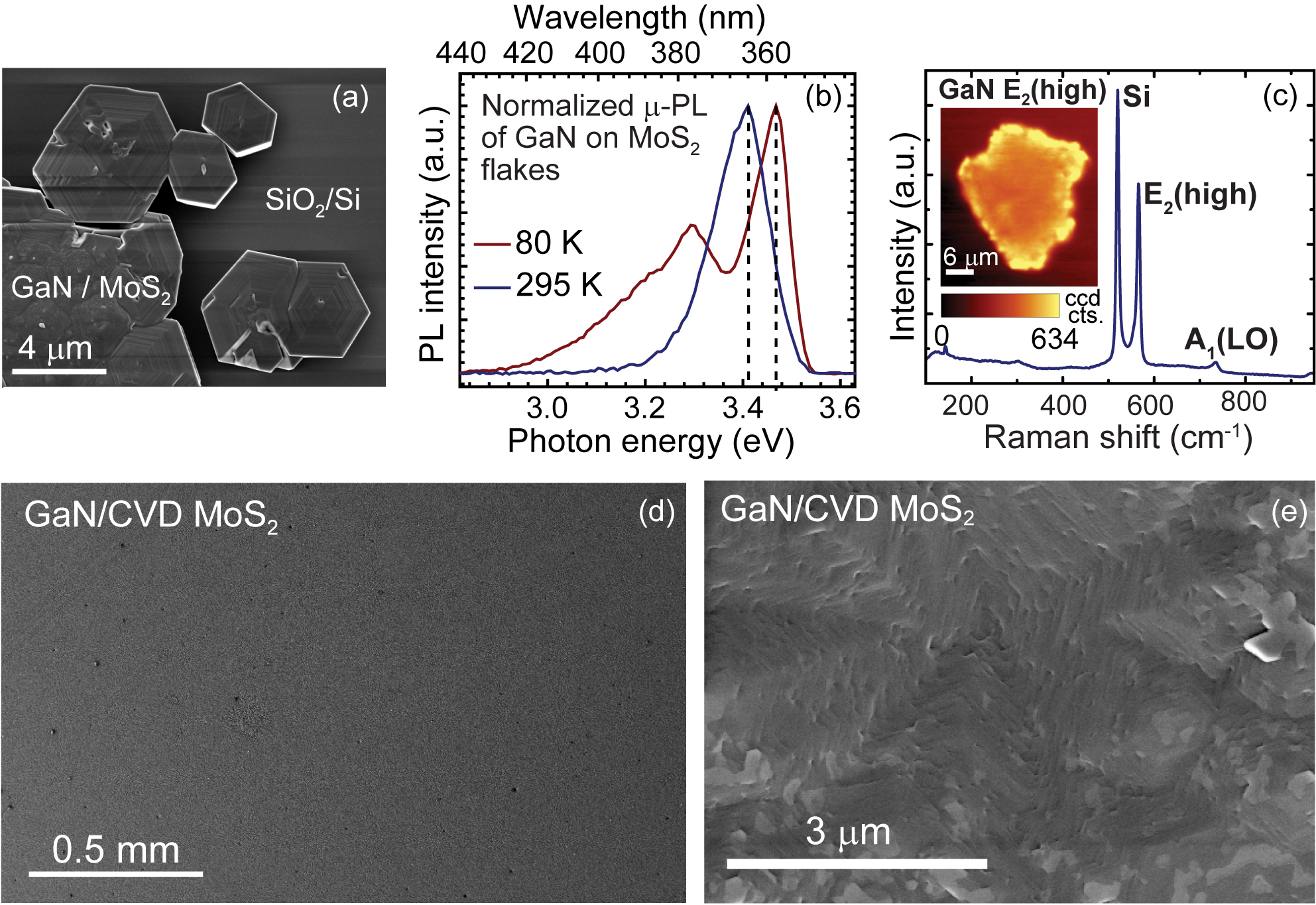}}
\caption{\onehalfspacing{\textbf{Growth of GaN on MoS$_2$.} (a) Hexagonal crystals of GaN which are obtained only on MoS$_2$ flakes (b) Room temperature $\mu$ - photoluminescence spectra showing strong near-band-edge emission from the GaN  (c) Spatially averaged Raman spectrum over GaN/MoS$_2$ flake (inset shows the integrated Raman map for the intensity of E$_2$(high) phonon mode of GaN). (d)and (e) SEM images showing the extension of GaN growth to large area CVD MoS$_2$. }}
\label{fig3}
\end{figure*}

Following a similar recipe, GaN epilayers were grown on MoS$_2$ as well. While one group had reported the molecular beam epitaxial growth of GaN on bulk MoS$_2$ in the 1990s,\cite{yamada1999molecular,yamada1999layered} there has been no systematic study. GaN layers were deposited on exfoliated MoS$_2$ using MOVPE and again nearly hexagonal crystals of GaN are obtained only on the MoS$_2$ flakes. A typical SEM image of the surface is shown in Figure~\ref{fig3}(a).
The GaN layer shows strong near-band-edge luminescence in room- and low-temperature photoluminescence (PL) (Figure~\ref{fig3}(b)). The details of photoluminescence are discussed later. Once again, the peak shift of the E$_2$(high) Raman mode is very small indicating nearly strain-free GaN on MoS$_2$ (Figure~\ref{fig3}(c)). The EBSD map of GaN grown on exfoliated MoS$_2$ (supplementary information section III) also shows that epitaxial GaN layer is single crystal similar to GaN grown on WS$_2$ and oriented in (0002) direction. From the low magnification and a zoomed-in SEM image of  GaN grown on large area CVD MoS$_2$ (Figure~\ref{fig3}(d) and \ref{fig3}(e)), it is clear that there is conformal coverage across the substrate and this method is scalable to larger substrate sizes.

\begin{figure}[ht]
\centerline{\includegraphics[width=88mm]{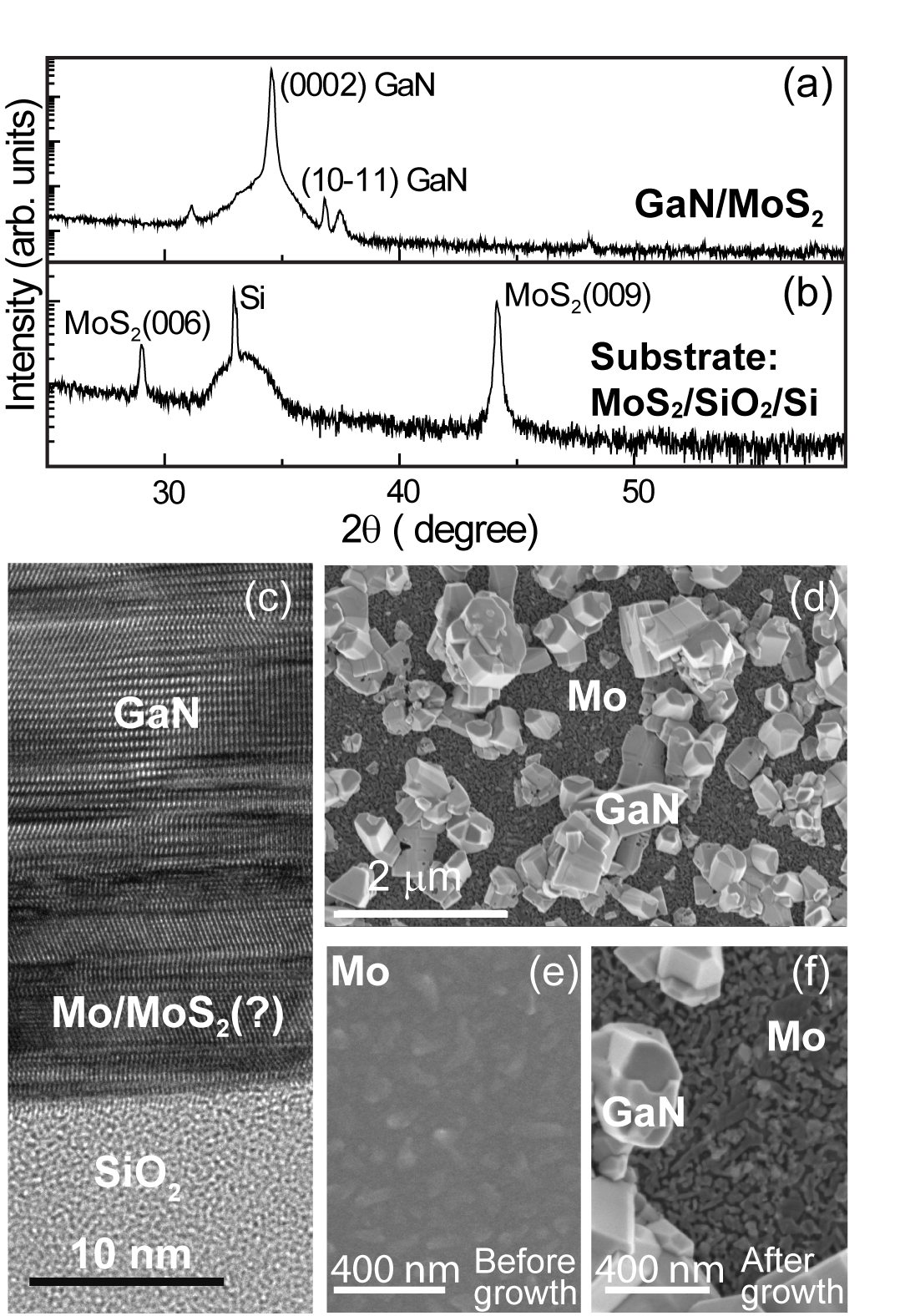}}
\caption{\onehalfspacing{\textbf{Necessity of MoS$_2$ for GaN growth.} (a) X-ray diffraction profile of GaN on MoS$_2$ shows preferential (0002) orientation but no MoS$_2$ peak after growth as compared to substrate XRD profile(b). (c) Cross-sectional transmission electron micrograph does not show MoS$_2$ at the interface of substrate and GaN epilayer. (d) The SEM shows faceted chunks of GaN on Mo substrate and no conformal coverage of GaN. The micrographs below (d) show sputtered Mo on sapphire before (d) and after (e) growth.}}
\label{fig4}
\end{figure}

While the GaN epilayers grown on exfoliated MoS$_2$ are preferentially oriented in (0002) direction as seen from XRD pattern (Figure~\ref{fig4}(a)), the surprising observation is that there are no characteristic MoS$_2$ peaks after the MOVPE growth (Figure~\ref{fig4}(a) and (b)). This suggests that the MoS$_2$ may have degraded under the MOVPE growth conditions (high growth temperature and exposure to ammonia and hydrogen). TEM imaging was used to confirm this hypothesis. As seen in Figure~\ref{fig4}(c), clear lattice fringes of the GaN layer are present in the top of the image. However, below the GaN, instead of the characteristic signature of layered MoS$_2$, only metallic molybdenum (Mo) was detected. A simple experiment to check the role of MoS$_2$ in the growth of GaN was to attempt to grow GaN on Mo metal alone using sputtered Mo layers deposited on sapphire. As seen in Figure~\ref{fig4}(d), faceted chunks of GaN were obtained and there was no conformal growth of GaN on Mo. The morphology of the the Mo surface also changed after the MOVPE growth of GaN on Mo (Figure~\ref{fig4}(e) and \ref{fig4}(f)). This suggests that the MoS$_2$ was indeed present and served as a substrate for the initial growth of the GaN epilayer, but degraded at some point during the high temperature growth.

\subsection{\normalsize{Comparative photoluminescence spectroscopy of GaN grown on WS$_2$ and MoS$_2$}}

\begin{figure*}[p]
\centerline{\includegraphics[width=1.0\textwidth]{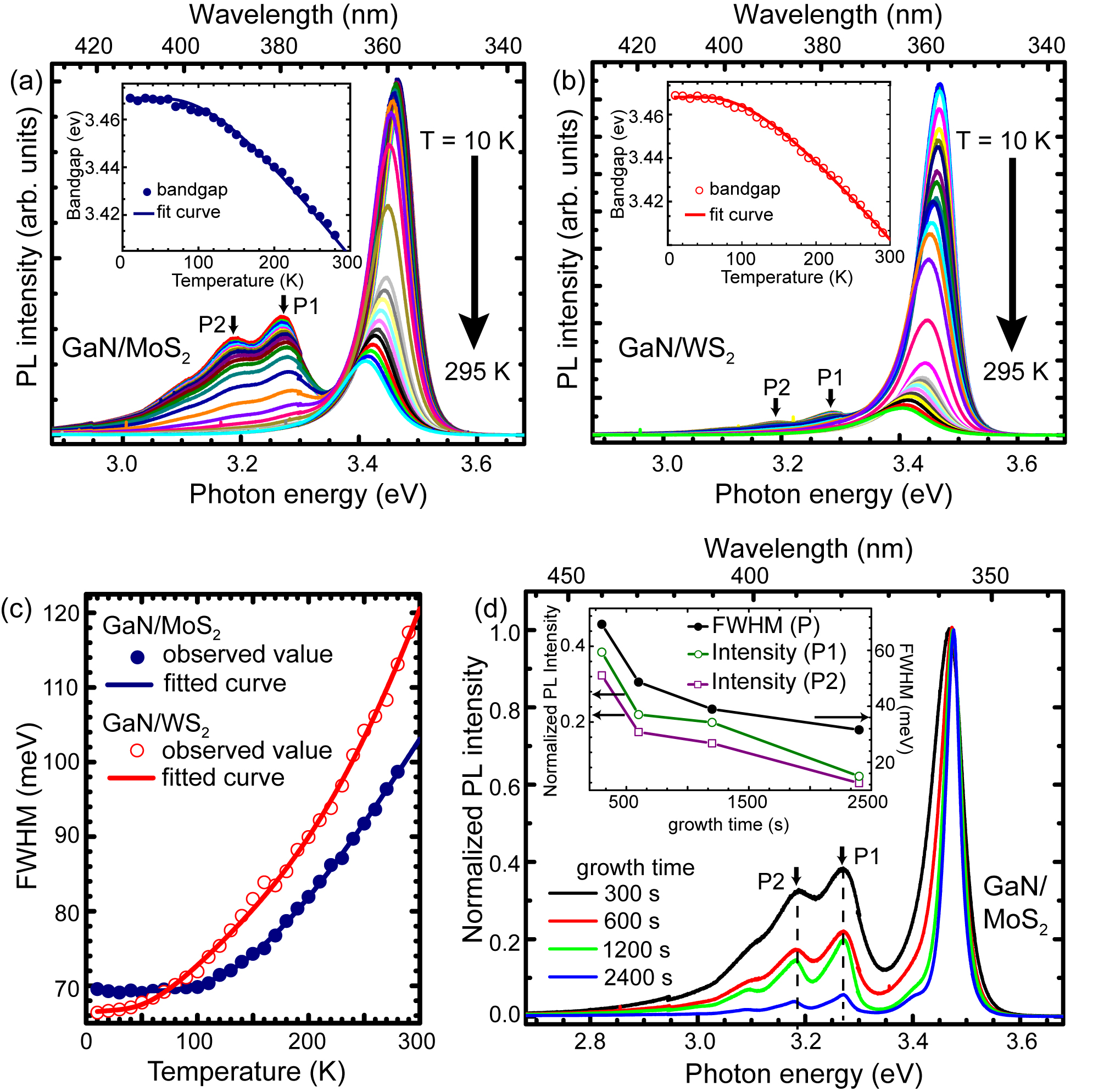}}
\caption{\onehalfspacing{\textbf{Comparative PL spectroscopy of GaN grown on WS$_2$ and on MoS$_2$.}  Temperature dependent photoluminescence of GaN grown (a) on WS$_2$  and (b) on MoS$_2$. Insets of (a) and (b) show the Bose-Einstein expression fit for the variation of NBE peak positions with temperature. (c) Temperature dependence of FWHM of NBE emission line for GaN/WS$_2$ and GaN/MoS$_2$. (d) Low temperature (10 K) PL of GaN grown on MoS$_2$ with different t$_{GaN}$ (inset shows that intensity of peaks P1 and P2 decreases  and NBE emission linewidth decreases with increasing t$_{GaN}$). }}
\label{fig5}
\end{figure*}

Temperature-dependent photoluminescence measurements were done on both the GaN/WS$_2$ and GaN/MoS$_2$ samples as shown in Figures~\ref{fig5}(a) and (b). The PL spectra exhibit a dominant near-band-edge (NBE) transition band at 3.41 eV (temperature (T) = 290~K), which clearly blue-shifts with decreasing temperature. Apart from this, there are two additional peaks at lower energies at 3.27 eV and 3.18 eV which we label as P1 and P2, respectively. We first discuss the behaviour of the NBE emission. The insets of Figures~\ref{fig5}(a) and \ref{fig5}(b) show the variation of NBE peak position with temperature which are fitted by the Bose-Einstein expression:\cite{lautenschlager1987interband}

\begin{equation}
\rm E(T) = E(0)-2a_B/[exp(\theta/T)-1] \label{first equation}
\end{equation}

where E(0) is the transition energy at T = 0 K and a$_B$ is the strength of the average exciton-phonon interaction and $\theta$ is the average phonon frequency. The values of E(T), a$_B$ and $\theta$ as obtained from the fitted curves (with 95~\% confidence bounds) are  shown in Table~1 and are in close agreement with the reported values for this temperature range.\cite{li1997temperature} The E(0) values obtained for the GaN grown on WS$_2$ and MoS$_2$ are  3.468$\pm$0.002~eV and  3.469$\pm$0.002~eV, respectively, very close to 3.471~eV\cite{tchounkeu1996optical} for unstrained GaN, again indicating strain-free layers.

 The FWHM of the NBE emission line increases with increasing temperature due to increasing exciton-phonon interaction at higher temperatures. The temperature dependence of the linewidth has the usual form:\cite{rudin1990temperature}

\begin{equation}
\Gamma (T) = \Gamma_{\rm inh} + \gamma_{\rm LA} T +
\dfrac{\Gamma_{\rm LO}}{\exp(\hbar \omega_{\rm LO}/k_{\rm B} T)-1} +
\Gamma_{\rm i}\exp\left(-\dfrac{E_{\rm i}}{k_{\rm B}T}\right)
\end{equation}

where $\Gamma_{\rm inh}$ is the inhomogeneous broadening term, $\gamma_{\rm LA}$ is a coefficient of exciton-acoustic-phonon interaction, $\Gamma_{\rm LO}$ is the exciton-LO-phonon coupling constant, $\omega_{\rm LO}$ is the LO-phonon energy, $\Gamma_{\rm i}$ is a proportionality factor which accounts for the concentration of impurity centers and $E_{\rm i}$ is the binding energy of impurity-bound excitons averaged over all possible locations of the impurities. The obtained value of $\Gamma_{\rm i}$  from the fit curves for WS$_2$ and MoS$_2$ (Figure~\ref{fig5}(c)) are 55$\pm$16~meV  and 171$\pm$87~meV, respectively which indicates that the concentration of impurity centers in GaN grown on WS$_2$ is lesser than that for the GaN grown on MoS$_2$. The same is also evident from the value of $E_{\rm i}$ which is lesser in case of  GaN/WS$_2$ compared to that for GaN/MoS$_2$ (Table 1).

\begin{table}[!h]
\caption{The parameters E(0), a$_B$, $\theta$, $\Gamma_{\rm i}$ and $E_{\rm i}$ of GaN epilayers grown on WS$_2$ and MoS$_2$ (with 95~\% confidence bounds)}
\begin{center}
\begin{tabular}{c | c c c c c}
\hline
GaN/substrate & E(0) & a$_B$ & $\theta$ & $\Gamma_{\rm i}$ & $E_{\rm i}$\\
& (eV) & (meV)& (K) & (meV) & (meV)\\
\hline
GaN/WS$_2$ & 3.468$\pm$0.002 & 78$\pm$18 & 365$\pm$48 &  55$\pm$16 & 20$\pm$5 \\
GaN/MoS$_2$ & 3.469$\pm$0.002 & 87$\pm$24 & 385$\pm$56 &  171$\pm$87 & 46$\pm$9\\
  \hline
\end{tabular}
\end{center}
\end{table}

 The peak intensity of P1 and P2 is larger in GaN/MoS$_2$ epilayer compared to GaN/WS$_2$ epilayer, which strongly suggests that these peaks arise from the defects originating from the degradation of the substrate. To confirm this hypothesis, GaN epilayers  of different thicknesses were grown on MoS$_2$ (t$_{GaN}$ = 300~s, 600~s, 1200~s and 2400~s). The low temperature (T~=~10~K) PL spectra for these samples, normalized to the NBE peak, are shown in Figure~\ref{fig5}(d). It can be seen that on increasing t$_{GaN}$,  i.e. with larger thickness of the GaN layer, the intensity of P1 and P2 decreases (inset of Figure~\ref{fig5}(d)). At our laser excitation energy  the absorption length in GaN is only $\sim$50~-~60~nm, hence the PL observed is mainly from the top $\sim$~200~nm GaN epilayer. The decrease in intensity of P1 and P2 thus indicates that the defect density in the top GaN layer reduces as the thickness increases. While this could be attributed to an overall reduction in extended defects, we believe that in our samples, P1 and P2 are related to point defects that originate from the GaN/substrate interface due to MoS$_2$ degradation. This was confirmed from the secondary ion mass spectrometry (SIMS) profile which showed a significant concentration of sulphur in the GaN epilayer (supplementary information section IV). Also, the FWHM of GaN/MoS$_2$ decreases on increasing t$_{GaN}$ (the inset of Figure~\ref{fig5}(d)), indicating better GaN quality  on increasing thickness. However, we have not found luminescence of sulphur impurities in GaN reported in literature.

\subsection{\normalsize{GaN grown on other TMDCs}}
GaN growth on other mechanically-exfoliated TMDCs like WSe$_2$, MoSe$_2$, ReS$_2$ and ReSe$_2$ (lattice mismatch to GaN -- $\sim$ 1~-~3 \%) was also attempted (SEM images in supplementary information section V). The micrographs clearly indicate that epitaxial growth of GaN is possible on these substrates. However the heat-up and the growth initiation steps would need to be optimized for the different TMDCs keeping in mind their thermal stabilities and reactivities. With the proper optimization of MOVPE growth conditions, these TMDCs can also be potential substrates for III-nitride growth.

\section{\large{Conclusion}}

We report the MOVPE growth of strain-free, single-crystal islands of GaN on mechanically-exfoliated flakes of WS$_2$ and MoS$_2$, discussing their structural and optical properties. We also present a detailed comparison of temperature-dependent PL of GaN grown on WS$_2$ and MoS$_2$ and a preliminary demonstration of large-area epitaxial growth of GaN on CVD MoS$_2$. Our investigations establish TMDCs as interesting near-lattice-matched substrates for GaN. With appropriate choice of substrates and growth conditions, it opens up the prospect of combining the III-nitrides with the transition metal dichalcogenides to realize novel heterostructures such as stacked layered MoS$_2$ and nitrides for solar energy conversion, as theoretically predicted\cite{zhang2014novel} recently. Further, growth on large-area single crystal TMDCs may provide a route to large-area single crystal GaN layers which could be released to serve as bulk substrates.

\section{{\large{Experimental section}}}

\small{\textbf{Preparation of substrates.} All the TMDCs were synthesized by first reacting the constituent elements in stoichiometric ratio to form precursors followed by iodine vapour transport in order to get bulk crystals suitable for exfoliation into thin films (details in supplementary information section II). In the case of MoS$_2$, we used naturally available bulk MoS$_2$ crystals for exfoliation. Raman spectroscopy details of the exfoliated and CVD grown MoS$_2$, and exfoliated WS$_2$ substrate materials are provided in supplementary information section VI.}

\small{\textbf{Growth of GaN epilayer.} GaN layers were deposited on these TMDC substrates using low pressure MOVPE in a 3x2" close-coupled showerhead system using standard trimethylgallium (TMGa) and NH$_3$ precursors.  H$_2$ carrier gas was used for the GaN layer growth with samples being heated up and cooled down using N$_2$ carrier gas. The growth rate of GaN on MoS$_2$ is $\sim$~50~-~60~nm per minute.}

\small{\textbf{Characterization techniques.} The films were structurally characterized using field-emission scanning electron microscopy (FE-SEM), X-ray diffraction (XRD), transmission electron microscopy (TEM) and electron back scattering diffraction (EBSD). The optical properties were measured by photoluminescence (PL) spectroscopy. Temperature-dependent PL measurements were done in the 10~K~-~295~K range using a set-up with a frequency-quadrupled 266~nm Nd:YAG laser for excitation, and a 0.55m monochromator equipped with a cooled Si-CCD detector. Confocal Raman spectroscopy measurements were performed on the samples using 532~nm laser excitation. The Raman peak shift and the full width half maxima (FWHM) are determined by fitting a Lorentzian function to the observed data. For comparing the peak positions, the spectrum is aligned with reference to the Si substrate peak (520 cm$^{-1}$).}


\section*{\small{Acknowledgements}}
\small{The work at TIFR was supported by the Government of India under project 12P0168. The authors are thankful to Sandip Ghosh for his valuable help and discussions in PL measurements, and acknowledge the support of B. A. Chalke and R. D. Bapat for help in EBSD measurements. We acknowledge Carina B. Maliakkal and Nirupam Hatui for assistance with materials characterization. We thank Mandar M. Deshmukh for providing bulk MoS$_2$ crystals for exfoliation.  }



\section*{Supplementary material}

**Layered transition metal dichalcogenides: promising near-lattice-matched substrates for GaN growth (Supplementary Information)**

Priti Gupta,[1, a)] A. A. Rahman,[1] Shruti Subramanian,[1, b)] Shalini Gupta,[1, 2] Arumugam Thamizhavel,[1] Tatyana Orlova,[3] Sergei Rouvimov,[3] Suresh Vishwanath,[3, c)] Vladimir Protasenko,[3] Masihhur R. Laskar,[4] Huili Grace Xing,[3, c)] Debdeep Jena,[3, c)] and Arnab Bhattacharya[1, d)]

[1)] *Department of Condensed Matter Physics and Materials Science, Tata Institute of Fundamental Research, Mumbai, India*
[2)] *UM-DAE Center for Excellence in Basic Sciences, Mumbai, India*
[3)] *Department of Electrical Engineering, University of Notre Dame, Notre Dame, USA*
[4)] *Department of Chemical and Biological Engineering, University of Wisconsin-Madison, Madison, USA*

(Dated: 28 September 2015)

---

[a)]Presently at Department of Materials Science and Metallurgy, University of Cambridge, Cambridge, UK
[b)]Presently at Department of Materials Science and Engineering, Pennsylvania State University, Pennsylvania, USA
[c)]Presently at Department of Electrical and Computer Engineering, Department of Materials Science and Engineering, Cornell University, Ithaca, NY, USA
[d)]Email: arnab@tifr.res.in

# I. LATTICE PARAMETERS AND BANDGAP VALUES OF TMDCS AND III-NITRIDES

The bandgap versus lattice parameter diagram (Figure 1 of the main manuscript) was plotted based on the room temperature bandgap values of the monolayers of the TMDC materials as available in the literature

**III-nitrides**

- **AlN:** M. Feneberg, M. F. Romero, B. Neuschl, K. Thonke, M. Röppischer, C. Cobet, N. Esser, M. Bickermann and R. Goldhahn, *Applied Physics Letters* **102** (2013), 052112.

- **GaN:** B. Monemar, *Physical Review B* **10** (1974), 676.

- **InN:** J. Wu, W. Walukiewicz, K. Yu, J. Ager Iii, E. Haller, H. Lu, W. J. Schaff, Y. Saito and Y. Nanishi, *Applied Physics Letters* **80** (2002), 3967–3969.

**TMDCs**

- **WS$_2$:** A. Berkdemir, H. R. Gutiérrez, A. R. Botello-Méndez, N. Perea-López, A. L. Elías, C.-I. Chia, B. Wang, V. H. Crespi, F. López-Urías, J.-C. Charlier *et al.*, *Scientific Reports* **3** (2013).

- **MoS$_2$, MoSe$_2$:** S. Tongay, J. Zhou, C. Ataca, K. Lo, T. S. Matthews, J. Li, J. C. Grossman and J. Wu, *Nano letters* **12** (2012), 5576–5580.

- **WSe$_2$:** P. Tonndorf, R. Schmidt, P. Böttger, X. Zhang, J. Börner, A. Liebig, M. Albrecht, C. Kloc, O. Gordan, D. R. Zahn *et al.*, *Optics express* **21** (2013), 4908–4916.

- **ReSe$_2$:** C. Ho, P. Liao, Y. Huang, T. Yang and K. Tiong, *Journal of Applied Physics* **81** (1997), 6380–6383.

- **ReS$_2$:** S. Tongay, H. Sahin, C. Ko, A. Luce, W. Fan, K. Liu, J. Zhou, Y.-S. Huang, C.-H. Ho, J. Yan *et al.*, *Nature Communications* **5** (2014).

- **HfS$_2$:** D. L. Greenaway and R. Nitsche, *Journal of Physics and Chemistry of Solids* **26** (1965), 1445–1458.

- **TiS$_2$:** Y.-H. Liu, S. H. Porter and J. E. Goldberger, *Journal of the American Chemical Society* **134** (2012), 5044–5047.

- **ZrS$_2$:** M. Moustafa, T. Zandt, C. Janowitz and R. Manzke, *Physical Review B* **80** (2009), 035206.

The lattice parameters were obtained from their respective powder diffraction file in the database of Joint Committee on Powder Diffraction Standards (JCPDS), International Center for Diffraction Data, Newtown Square, PA (2013). The following is the list of JCPDS card numbers in the database:

**III-nitrides**

- **AlN:** 00-025-1133
- **GaN:** 00-050-0792
- **InN:** 00-050-1239

**TMDCs**

- **WS$_2$:** 04-003-4478
- **WSe$_2$:** 00-038-1388
- **MoS$_2$:** 00-037-1492
- **MoSe$_2$:** 04-004-8782
- **ReS$_2$:** 04-002-2231
- **ReSe$_2$:** 00-050-0537
- **HfS$_2$:** 00-028-0444
- **ZrS$_2$:** 00-011-0679
- **TiS$_2$:** 01-070-6204

## II. PREPARATION OF TRANSITION METAL DICHALCOGENIDES (TMDC) SUBSTRATES

Transition metal dichalcogenides (TMDC) of the type MX$_2$ where M is a transition metal and X is a chalcogen, are 2D layered materials, with van der Waals forces coupling the layers[?][?]. They can thus be easily exfoliated into thin monolayer to few-layer sheets, similar to graphene. We first prepared the TMDC crystals by the direct synthesis from their constituent elements. These crystals were exfoliated to get flakes that were transferred to a SiO$_2$/Si wafer as substrates for GaN growth.

### A. Crystal growth of TMDCs

A two-step process was followed for the crystal growth of the MX$_2$ (M= Mo, W, Re and X= S, Se). First the precursors for the crystal growth were prepared by the usual solid-state

reaction of a stoichiometric mixture of the respective high purity constituent elements in an evacuated and sealed quartz ampoule ($\sim 10^{-5}$ torr). This quartz ampoule was placed inside a carbolite box furnace, which was slowly ramped to the reaction temperature at the rate of 15 - 20 $^oC$/hr and kept at this temperature (1000 - 1200 $^oC$) for 3 - 4 days , then cooled down to room temperature at 50 - 60 $^oC$/hr. This resulted in flakes of the chalcogenide materials, typically 10 - 100 $\mu$m in size being synthesized. This product was then used in a second step to grow single crystals by the iodine vapor transport method. The reacted product ($\sim$ 1 g) along with $I_2$ (150 mg) were inserted in a 20 cm long, cleaned quartz tube, sealed and placed inside a 2-zone furnace. An appropriate temperature gradient ($\sim$ 100 - 120 $^oC$) was maintained across the tube for a period of 4-8 days to enable the vapor transport and recrystallization of the material. Temperature parameters and duration during iodine vapor transport of different TMDCs are listed in Table 1.

TABLE I. Temperature parameters and duration during iodine vapor transport of different TMDCs

| Initial material | Temperature Distribution | | Growth Time (Days) |
|---|---|---|---|
| | Hot Zone (°C) | Cold Zone (°C) | |
| $WS_2$ | 950 | 850 | 7 |
| $MoS_2$ | 1000 | 875 | 7 |
| $ReS_2$ | 1020 | 975 | 8 |
| $WSe_2$ | 800 | 700 | 4 |
| $MoSe_2$ | 1080 | 1020 | 5 |
| $ReSe_2$ | 1020 | 975 | 8 |

### B. Exfoliation of TMDCs crystals

The TMDCs crystals obtained are then exfoliated using 3M scotch tape and stamped on 300 nm thick $SiO_2$-coated Si ($\sim$ 1 $cm^2$). This is then used as a substrate for the epitaxial growth of GaN. Since our aim was to check the feasibility of the TMDC as a substrate for growth, we did not necessarily optimize the exfoliation to get monolayer sheets.

### C. CVD growth of $MoS_2$

Large area $MoS_2$ films were synthesized by the vapor phase sulphurization of thin Mo films deposited on sapphire substrates[?] . Typically $\sim$ 10 - 15 nm thick Mo layers were allowed to react with sulfur in an evacuated and sealed quartz tube at 900-1000 $^oC$ for 1 hour. Mo films deposited by sputtering and electron beam evaporation were both used with no significant difference in the quality of the resultant $MoS_2$.

## III. ELECTRON BACK SCATTER DIFFRACTION (EBSD) OF GaN/MoS$_2$

The EBSD map of GaN grown on exfoliated MoS$_2$ (Figure 1) clearly shows that epitaxial GaN layer is single crystal similar to GaN grown on WS$_2$ and oriented in (0002) direction.

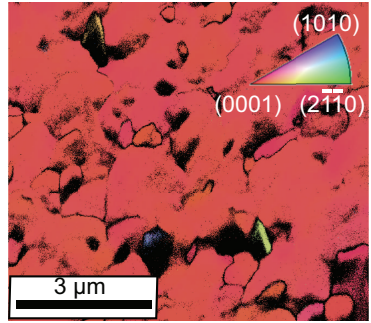

FIG. 1. **EBSD of GaN grown on exfoliated MoS$_2$** The map shows that GaN epilayer is single crystal and oriented in (0002) direction

## IV. SECONDARY ION MASS SPECTROMETRY (SIMS) OF GaN GROWN ON CVD MoS$_2$

As discussed in the main manuscript, MoS$_2$ was degrading after MOVPE growth of GaN. From an analysis of the behaviour of the photoluminescence of GaN on MoS$_2$ layers, we surmised that the degradation of MoS$_2$ left behind sulphur impurities in the layer. This was confirmed from the SIMS profile (done at Evans Analytical Group Inc.) which clearly shows a significant presence of sulphur in the GaN epilayer (Figure 2).

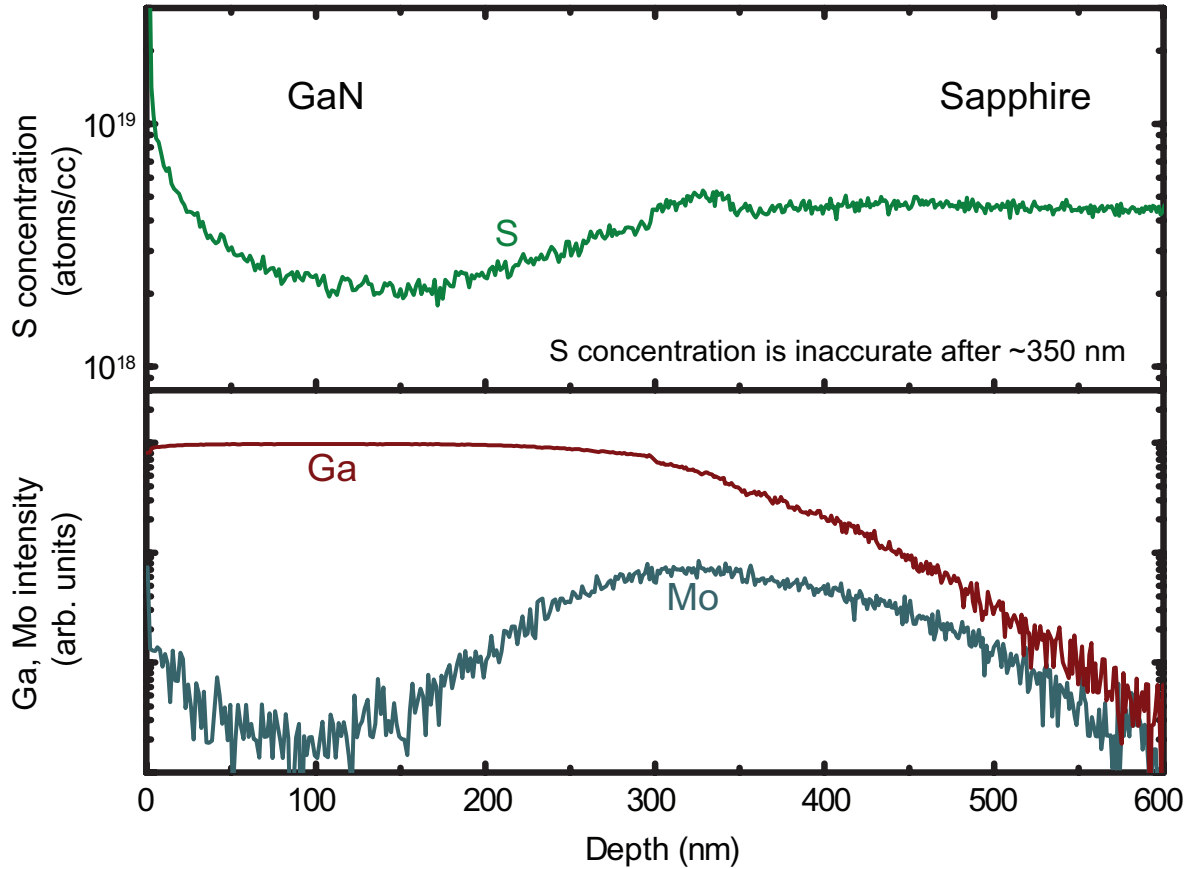

FIG. 2. **SIMS profile of GaN grown on CVD MoS$_2$/sapphire.** The profile shows significant presence of sulphur in the GaN epilayer.

6

## V. GaN GROWN ON OTHER TMDCs

Growth of GaN epilayers on other exfoliated TMDCs like $WSe_2$, $MoSe_2$, $ReS_2$ and $ReSe_2$ was also attempted and Figure 3 shows the corresponding SEM images. All these samples were grown following the same recipe used for the growth of GaN on $WS_2$ — a short initial layer at 900 ºC, followed by 300 s growth at 1040 ºC. From the micrographs, it is clear that epitaxial growth of GaN is possible on these substrates. However keeping in mind the different thermal stabilities of the various TMDCs, these conditions would need to be optimized independently for each material.

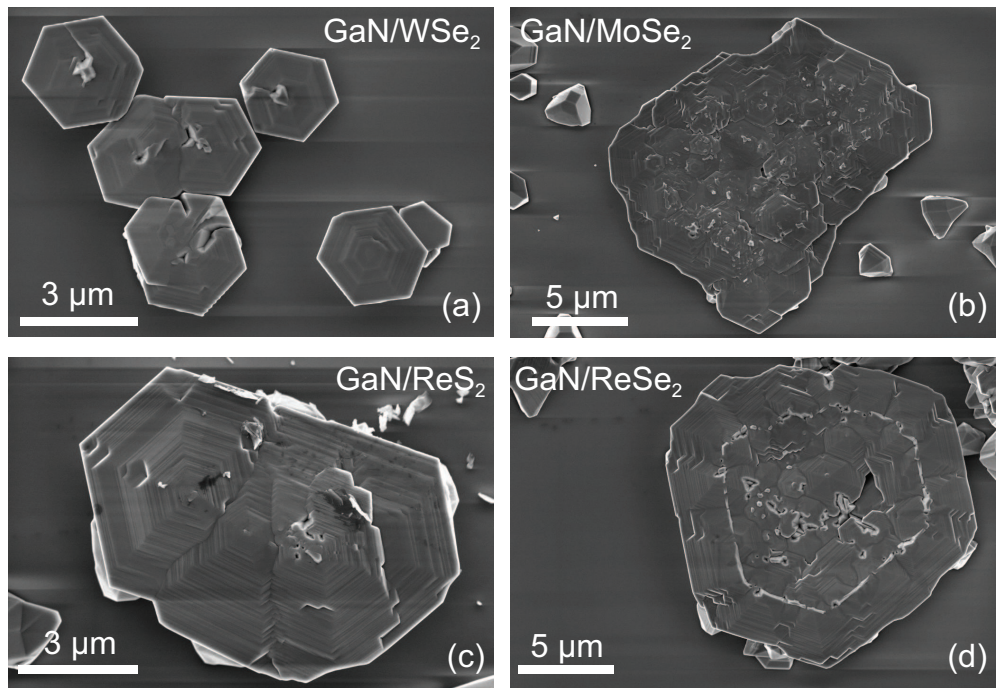

FIG. 3. **GaN grown on other TMDCs.** SEM images showing GaN grown on (a) $WSe_2$ (b) $MoSe_2$ (c) $ReS_2$ and (d) $ReSe_2$.

## VI. RAMAN SPECTROSCOPY MEASUREMENT OF $MoS_2$ AND $WS_2$ SUBSTRATES

Figure 4 shows the Raman spectra of the bare substrates, which clearly shows two prominent peaks $E^1_{2g}$ and $A_{1g}$. The peak positions of these modes indicate the number of layers in the TMDCs[?,?]. The peak frequencies of $E^1_{2g}$ and $A_{1g}$ modes correspond to bulk $MoS_2$, both in case of exfoliated and CVD $MoS_2$. While in case of $WS_2$ which is prepared by exfoliating synthesized $WS_2$ bulk crystal, the number of layers varies from 3 to bulk all over the sample.

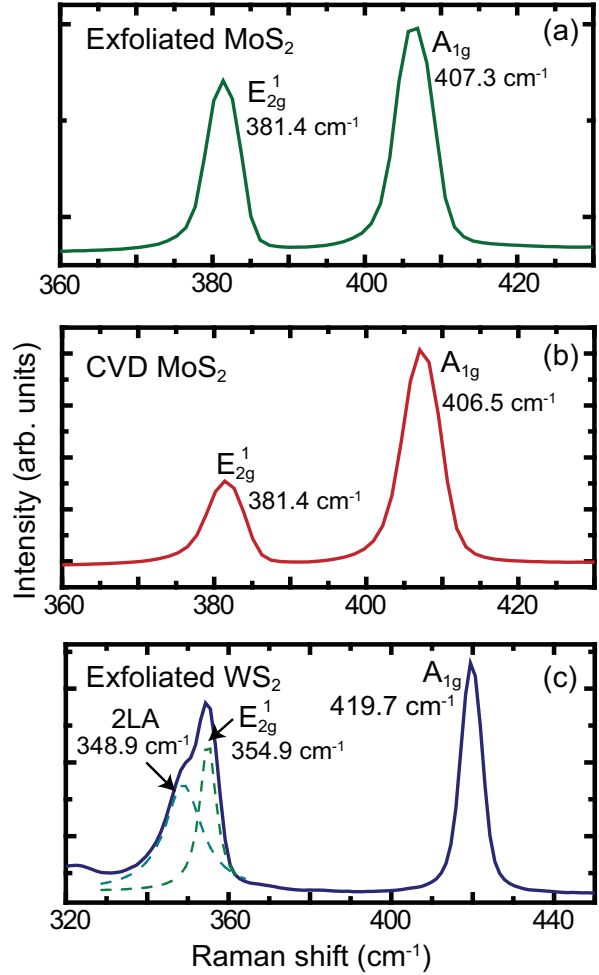

FIG. 4. **Raman spectra of bare substrates** (a) Exfoliated $MoS_2$ (b) CVD $MoS_2$ and (c) exfoliated $WS_2$ (dashed curves shows the Lorentzian two-peak fit for $A_{2g}$ and 2LA modes).



\end{document}